# Temperature-dependent steady and transient emission properties of InGaN/GaN multiple quantum well nanorods

Bin Jiang,[1] Chunfeng Zhang,[1*] Xiaoyong Wang,[1] Fei Xue,[1] Min Joo Park,[2] Joon Seop Kwak,[2*] Min Xiao[1, 3]

[1] National Laboratory of Solid State Microstructures and Department of Physics, Nanjing University, Nanjing 210093, China

[2] Department of Printed Electronic Engineering, Sunchon National University, Sunchon, Jeonnam 540-742, Korea

[3] Department of Physics, University of Arkansas, Fayetteville, Arkansas 72701, USA

*cfzhang@nju.edu.cn, jskwak@sunchon.ac.kr

Abstract：

We observed different temperature-dependent behaviors of steady and transient emission properties in dry-etched InGaN/GaN multiple-quantum-well (MQW) nanorods and the parent MQWs. To clarify the impacts of nanofabrication on the emission properties, time-resolved photoluminescence spectra were recorded at various temperatures with carrier density in different regimes. The confinement of carrier transport was observed to play an important role to the emission properties in nanorods, inducing different temperature-dependent photoluminescence decay rates between the nanorods and MQWs. Moreover, together with other effects, such as surface damages and partial relaxation of the strain, the confinement effect causes faster recombination of carriers in nanorods.

Recently, nano-texturing has been widely performed on InGaN/GaN multiple-quantum-wells (MQWs) to optimize the light-emitting-diode (LED) devices.[1-7] Compared with the planar MQWs, InGaN/GaN MQW nano-LEDs show better performance with improved brightness, luminescent efficiency, and color tunability over the whole visible spectral range.[3-7] The impacts of nanofabrication, including the light extraction enhancement, the strain relaxation, and the ease of sample growth with high-indium concentration, have been regarded as major factors for these improvements.[1-10]

The physics underlying the superior emission properties in InGaN/GaN heterostructures are under extensive study. The recombination of localized states is generally accepted as an origin of the high luminescent efficiency.[11-16] This recombination pathway is efficient to compete with nonradiative recombination channels induced by the threading dislocations, strain induced quantum confined Stark effect (QCSE), and other effects.[17-20] In nanofabricated samples, faster carrier recombination has been observed as a result of strain relaxation and the surface state trapping.[2] Besides the above effects, other impacts of nanofabrication may also modify the emission properties. For instance, the dimensions of nano-LEDs, in the scale of $10^2$ nanometers, are close or even smaller than the carrier diffusion length.[21-23] This will result in the confinement of the carrier transport in InGaN/GaN nanostructures.[11,24] In this work, we investigated the nanofabrication impacts on the emission properties in InGaN/GaN nanorods with emphasis on the effect of carrier transport confinement. We carried out time-integrated (TI) and time-resolved (TR) photoluminescence (PL) experiments on dry-etched MQW nanorod LED structures and the parent MQWs at various temperatures with a broad range of carrier density. We observed different temperature-dependent behaviors of steady and transient emission in nanorods and MQWs. While the emission intensity of TIPL in nanorods decreases monotonically with increasing temperature, the emission intensity in MQWs remains nearly unchanged with temperature up to 120 K. The thermally-activated carrier transport exhibits delayed-rise in TRPL traces recorded from MQWs when temperature increases,[11,24] but the delayed-rise characteristic is not present in the TRPL traces recorded from nanorods. This difference of temporal evaluation has been explained by the carrier transport confinement in nanorods.

The LED samples used in this study were grown on sapphire substrates by metal-organic chemical vapor deposition. The InGaN/GaN MQW LED structures consist of a GaN buffer, an n-type GaN layer, five pairs of 2.5 nm-thick $In_{0.1}Ga_{0.9}N$ quantum wells sandwiched between 13 nm GaN barriers, a p-type of GaN layer, as well as a p-type capping layer. The InGaN/GaN NRs with average diameters of 200 nm were fabricated by inductively-coupled-plasma etching using a novel mask of self-assemble ITO-based nanodots.[25] Scanning electron image of the nanorods and transmission electron image of a single nanorod are shown in Fig. 1. Frequency doubled optical pulses generated from a Ti: Sapphire regenerative amplifier (Libra, Coherent, 1 kHz, 90 fs) are used for pumping. The pumping wavelength at 400 nm only excites the InGaN well layers with excitation flux in the range of 7-117 μJ/cm$^2$ (corresponding to carrier density of $8 \times 10^{16} - 3 \times 10^{18} cm^{-3}$). Time-integrated emission spectra were measured with a fiber spectrometer (USB 2000+, Ocean-Optics). The technique of optical Kerr gating was utilized to do TRPL studies with Kerr medium of 5 mm thick carbon disulfide cell. The temporal resolution is about 2 ps. The PMT (5784-20, Hamamatsu) was used to record the TRPL traces with band pass optical filters at 450 nm with 40 nm bandwidth. The emission spectra at different decay time were recorded with a 0.5 m spectrograph (Sp 2500i, Princeton Instruments) equipped a liquid nitrogen cooled charge coupled devices. A liquid helium cryostat (MicroCryostatHe, Oxford) was used to do temperature-dependent measurements from 5 K to 300 K.

TIPL spectra in nanorods show some common emission features of InGaN/GaN heterostructures [Fig. 2]. With increasing excitation density, the emission shifts to shorter wavelength band in the spectra recorded at room temperature [Fig. 2 (a)]. This blue-shift has been explained by the combination of band-tail filling effect and reversal QCSE with high density of photo-excited carriers.[18,26-29] With decreasing temperature, PL intensity increases dramatically with slight blue-shift of the emission band [Fig. 2 (b)]. Compared with the parent MQWs, emission wavelength is shorter [Fig. 2 (c)] and emission intensity is much higher in NRs [Fig. 2 (f)] under same experimental configurations (i.e., the same excitation power and temperature). Temperature-dependence of emission properties show diverges in two samples as compared in Figs. 2 (c) and 2 (d). With increasing temperature, the emission peak shifts to the red side monotonically in NRs [Fig. 2 (c)]. In MQWs, the emission peak shifts to the blue side with temperature above 150 K, but further increasing temperature leads to a red-shift of emission peak [Fig.2 (c)]. While the emission intensity in nanorods decreases monotonically with increasing temperature, the emission intensity in MQWs remains nearly unchanged with temperature up to 120 K [Fig. 2 (d)]. To understand the detailed mechanisms, we probe the carrier dynamic behaviors with TRPL spectra.

As shown in Fig. 3 (a), the TRPL traces recorded from the InGaN nanorods show highly non-exponential decay. The recombination rate is strongly dependent on the excitation densities. To interpret the nonexponential decay dynamics, several models have been proposed in the last decade including a bi-dimensional donor-acceptor like recombination model,[30] a charge-separated dark state model,[16] and a frequently-used phenomenological stretched exponential (SE) function.[31] For simplicity, we will use the SE function to describe the carrier dynamics in this work. The SE function in the form of $I(t) = I_0 e^{-(t/\tau_0)^\beta}$ is a good mathematic description of the TRPL traces. In the function, $\tau_0$ is the SE decay lifetime, and $\beta$ expresses the distribution of rates with values between 0 (broad distribution) and 1 (narrow distribution).

Figure 3 (b) and 3 (c) compare the best fitted parameters of the TRPL traces in the NRs and MQWs at room temperature with different excitation densities. The carrier recombination rate is generally faster in nanorods than in MQWs [Figs. 3 (b), (d) and (e)]. As discussed in previous reports, [2,32] the surface trapping effect may be a reason for the faster recombination since the etching process can cause surface damages in nanorods. Also, other effects induced during nanofabrication can also change the dynamical behaviors of the PL emission in nanorods. The QCSE associated to the strain relaxation after nano-texturing may also lead to the faster recombination in nanorods. The QCSE, induced by the strain-induced internal electrical field, causes the spatial separation of confined electrons and holes in InGaN/GaN heterostructures and slows down the carrier recombination.[18,29,33] In the nanorods, the strain may be partially relaxed[1,2,8,9] and less internal electrical field will be present, leading to the faster relaxation in nanorods,[2] which is consistent with the power-dependent TIPL measurement as shown in Fig. 2 (e). With increasing excitation density, blue-shift of the emission peak, which is frequently considered as a signature of reversal QCSE (free carrier screening), is observed in both MQWs and nanorods. However, smaller blue-shifts were observed in the nanorod samples. These results indicate that the strain-induced QCSE still exists in the nanorods, in agreement with a recent study,[34] but the strain may be weaker in comparison to the parent MQWs.

As shown in the Figs. 3 (b) and 3 (c), the carrier recombination dynamics shows strongly dependence on the excitation density. The recombination lifetime decreases rapidly with carrier density below a critical value but the change becomes smaller and converges to a value at high carrier density in both samples. The reversal QCSE and the band-tail filling effect are frequently assigned to the power-dependent variations of recombination rate. With increasing excitation power, the internal electrical field may be screened and compensated by photon-excited carriers, leading to the reversal QCSE with the observed decrease of carrier recombination lifetime.[18,29] This effect also exhibits the time-dependent descreening phenomena with red-shift of PL spectra over the decay time.[33] Initially, the high density carriers induced by the intense laser pulses compensate the internal field efficiently. The carrier density decreases over the delay time so that the QCSE becomes more important and the recombination slows down.[18,29,33] In Figs. 3 (d) and 3 (e), we compare the counter-plotted image of the intensity as function of delay time and emission wavelength in the nanorods and MQWs, respectively, with excitation density of 117 $\mu J/cm^2$. Less red-shift of the emission over time can be observed in nanorods, indicating less strain in nanofabricated samples.

Band tail filling effect may also cause the temperature-dependent variation of recombination rate.[11,24,26] In general, this effect should not cause the discrimination of carrier dynamics in nanorods and the parent MQWs, since the band structure change is neglectable in the nanorods with diameters much larger than the effective Bohr radius (~ 3 nm).[27] Nevertheless, considering the confinement of carrier transport, the filling of lower energy levels can lead to the different recombination rates between the nanorods and MQWs. In InGaN MQWs, it has been observed that excitation transfers from the weakly-localized states to the strongly-localized state with increasing temperature.[11,24] This thermally-activated carrier transport drives the excitons filling the low energy levels prior to the interband transition. The diameters of nanorods are close or smaller than the carrier diffusion length,[21-23] so that the carrier transport will be partially eliminated in nanorods.

The influence of carrier transport confinement on the emission properties in nanorods is evidenced by the transient spectral results. With relatively low carrier density, the temperature-dependent TRPL spectra recorded from the MQWs and nanorods are compared in Figs. 4 (a) and 4 (b). Instead of immediate decay post-excitation, slight rise of the emission over time can be distinct in the TRPL curves recorded from MQW samples. This delayed-rise becomes more explicit with increasing the temperature up to 160 K. It is observable till the temperature goes up to 250 K, above which the non-radiative decay process dominates. The delayed-rise originates from intra-well carrier transport from weakly-localized states (higher potential minima) to strongly-localized states (lower potential minima).[11,24] This temperature dependence indicates the carrier transport is activated thermally. The carrier transport process, in a time scale of sub-nanosecond,[21-23] enables that photon-excited electrons and holes to recombine from the strongly-localized states in MQWs. In nanorods, the size effect has profound impacts on carrier transport process and the temporal evolution behavior of the emission in nanorods diverges from that in MWQs. As shown in Fig. 4 (b), with increase of the temperature, the delayed-rise is not observable in the TRPL traces in nanorods. This difference is a clear signature that the thermally-activated carrier transport process[11,24] has been strongly modified and partially eliminated by the nanofabrication processes.

For providing an intuitive description, a schematic diagram in Fig.4 (c) is used to illustrate the effect of carrier transport confinement on the emission dynamics in nanorods. Due to the composition fluctuation, the potential minima (the solid line) distribute differently in space in the planner MQWs. Generally, the recombination lifetime from shallower potential minima (weakly-localized states) is faster than that from deeper potential minima (strongly-localized states).[11,24,32] At the low temperature (5 K), thermal energy is insufficient for carriers to conquer the weak potential barriers. When the temperature increases, thermal energy will assist the carriers to "escape" from the weakly-localized states and transport to the strongly-localized states, which causes the delayed-rise in the TRPL curves. However, in the nanorod samples, the boundaries (the dashed line) will prevent the carrier transport process, so that the probability of carrier recombination in the nanorods from the weakly-localized states or the surface states becomes much higher than that in the MQWs. The confinement effect will lead to the absence of the delayed-rise and faster recombination in nanorods as one of the major factors that cause the different emission dynamics in the InGaN nanorods and MQWs.

Hence, multiple mechanisms may be involved in the different emission dynamics observed in nanorods and MQWs with faster PL recombination and different temperature-dependent temporal evolutions in nanorods. Although, many effects including the surface damages, the partial relaxation of strain, and the carrier transport confinement could all contribute to the faster recombination in nanorods, the discrimination of temperature dependence, characterized by the "delayed-rise" in the TRPL traces, should be mainly attributed to the carrier transport confinement.

With the above discussions on carrier dynamics, we can get a better understanding of the mechanisms on TI emission properties in nanorods. The nanofabrication influences the carrier recombination processes and causes the faster carrier recombination with a shorter emission wavelength in the nanorods,[2,11,24,26] resulting in different peak shifts of PL emission in the nanorods and MQWs with increasing excitation density or decreasing temperature. The confinement of the carrier transport processes in nanorods leads to different temperature dependences of the emission peak positions. With an increasing temperature, the thermally-activated carrier transport enables most of carriers to recombine from the strongly-localized states in MQWs. This carrier transport may compensate the increasing non-radiative recombination with temperature up to 120 K [Fig. 2 (d)]. In nanorods, the carrier transport process is partially eliminated and the emission intensity drops monotonically with increasing temperature. This causes a stronger temperature dependence of PL intensity in nanorods. The modification of the carrier transport in nanorods could cause relatively low quantum efficiency in nanorods at room temperature. Nevertheless, the light extraction efficiency is significantly improved.[2,5,6,10] Higher power of PL emission is observed in nanorods [Fig. 2 (f)], even though the number of photo-excited carriers is much less since the filling factor is about 50% in nanorods. Since the total output power is the ultimate goal of LED design, optimizing NR diameters is of promise to improve the device performance.

In conclusion, we have performed a systematic study on emission properties of InGaN/GaN nanorod LED structures by TRPL and TIPL experiments at different power densities and temperatures. We observed different temperature behaviors of steady and transient emission properties in nanorods and MQWs. These differences were explained by the strain relaxation effect and the confinement of carrier transport in the nanorods. The impacts of nanofabrication on carrier dynamics revealed in this study are key factors for the nano-LED design. Device performance can be improved with a tradeoff between internal quantum efficiency and light extraction efficiency in nanorods.

This work is supported by the Program of International S&T Cooperation (2011DFA01400, MOST), the National Basic Research Program of China (2012CB921801, and 2011CBA00205), National Science Foundation of China(61108001, and 11021403), NRF of Korea grant funded by the Korea government (MEST) (K2011-0017325). The author C.Z. acknowledges financial support from New Century Excellent Talents program (NCET-09-0467), PAPD and Fundamental Research Funds for the Central Universities (1107020420, 1118020406, 1104020403, and 1115020404). The author J.S.K acknowledges financial support from WCU program in SCNU.

# Figure captions:

Figure 1: Scanning electron microscopy image of the morphologies of the InGaN/GaN nanorods. Inset shows the transmission electron microscopy image of the cross section of a single nanorod.

Figure 2: Emission properties of InGaN/GaN nanorods. (a) Room-temperature emission spectra of InGaN/GaN nanorods recorded with different excitation densities; (b) Emission spectra of InGaN/GaN nanorods recorded at different temperature with excitation density of 25 μJ/cm$^2$; (c) Photon energy of Emission peaks from nanorods and MQWs is plotted as functions of temperature with excitation densities of 25 μJ/cm$^2$ and 117 μJ/cm$^2$, respectively; (d) Intensities of emission from nanorods and MQWs are plotted as functions of temperature with excitation densities of 25 μJ/cm$^2$ and 117 μJ/cm$^2$, respectively. (e) and (f) plot the photon energy of emission peak and the emission intensity recorded in the samples of nanorods and MQWs as function of excitation density at 5 K and 300 K, respectively.

Figure 3: (a) Logarithmic plot of TRPL traces recorded from InGaN/GaN nanorods at room temperature with different excitation densities. The SE fitting parameters ($\tau_0$ and $\beta$) of the TRPL traces recorded from nanorods and MQWs are plotted as functions of excitation densities in (b) and (c). Logarithmic plot of emission intensity as a function of emission wavelength and decay time in parent MQWs (d) and NRs (e) with excitation density of 117 μJ/cm$^2$ at room temperature.

Figure 4: Temperature-dependent TRPL traces recorded from MQWs (a) and nanorods (b) with excitation density of 25 μJ/cm$^2$. The curves are vertically shifted for clarity. (c) Schematic diagram of the confinement of carrier transport by the nanorods. The solid line represents the spatial distribution of potential; The dashed line represents the nanorod side boundaries; The arrows represent the carrier removal processes of thermal escape, radiative and non-radiative recombinations, respectively.

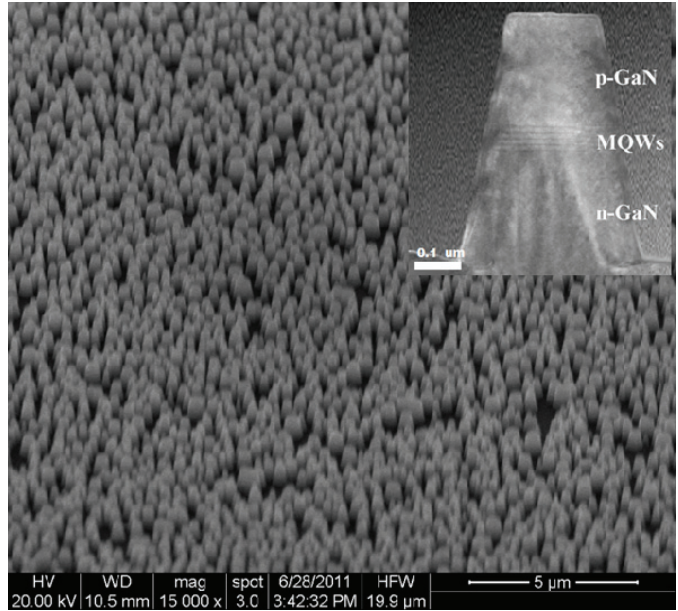

**Jiang et al., Figure 1**

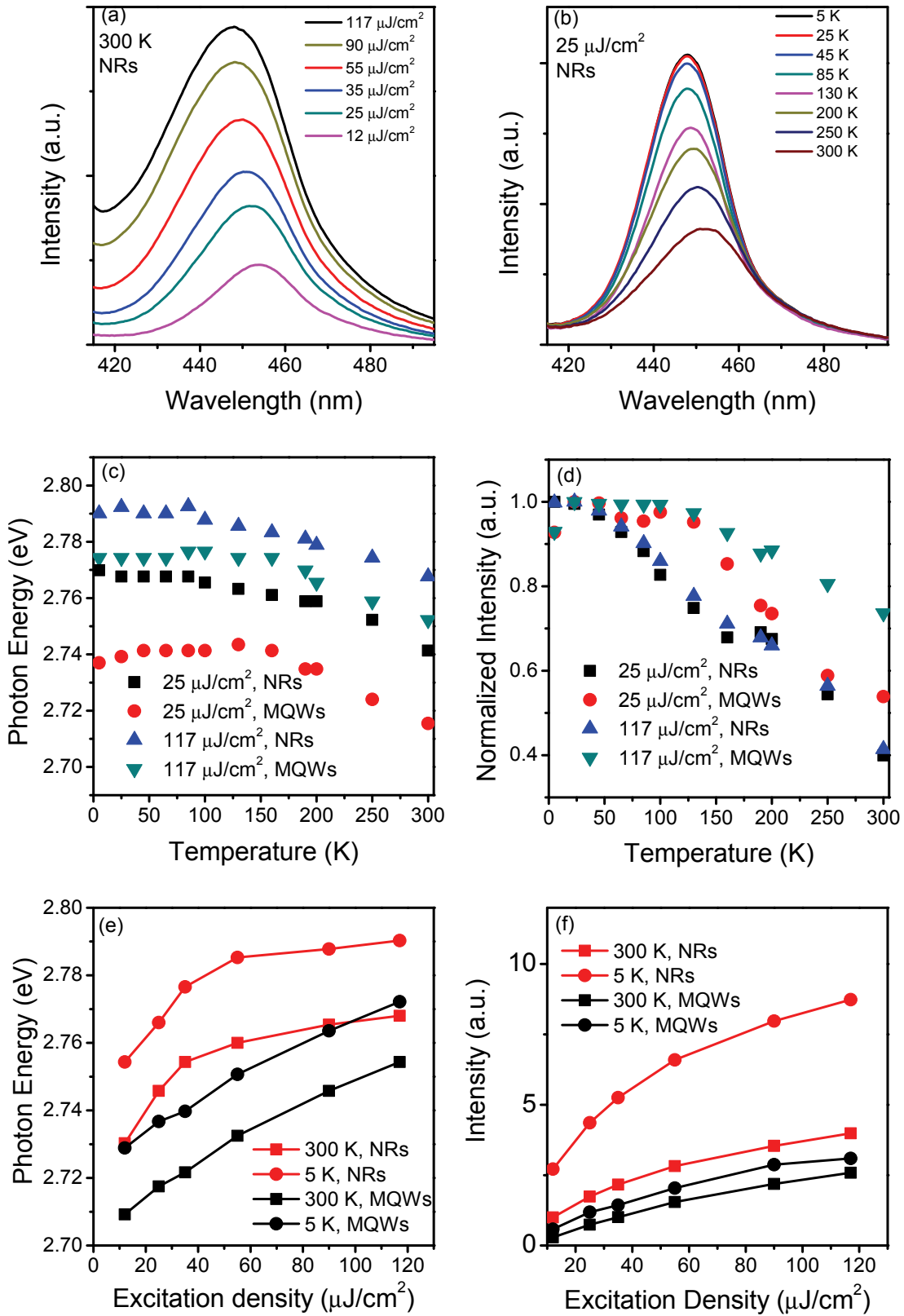

Jiang et al., Figure 2

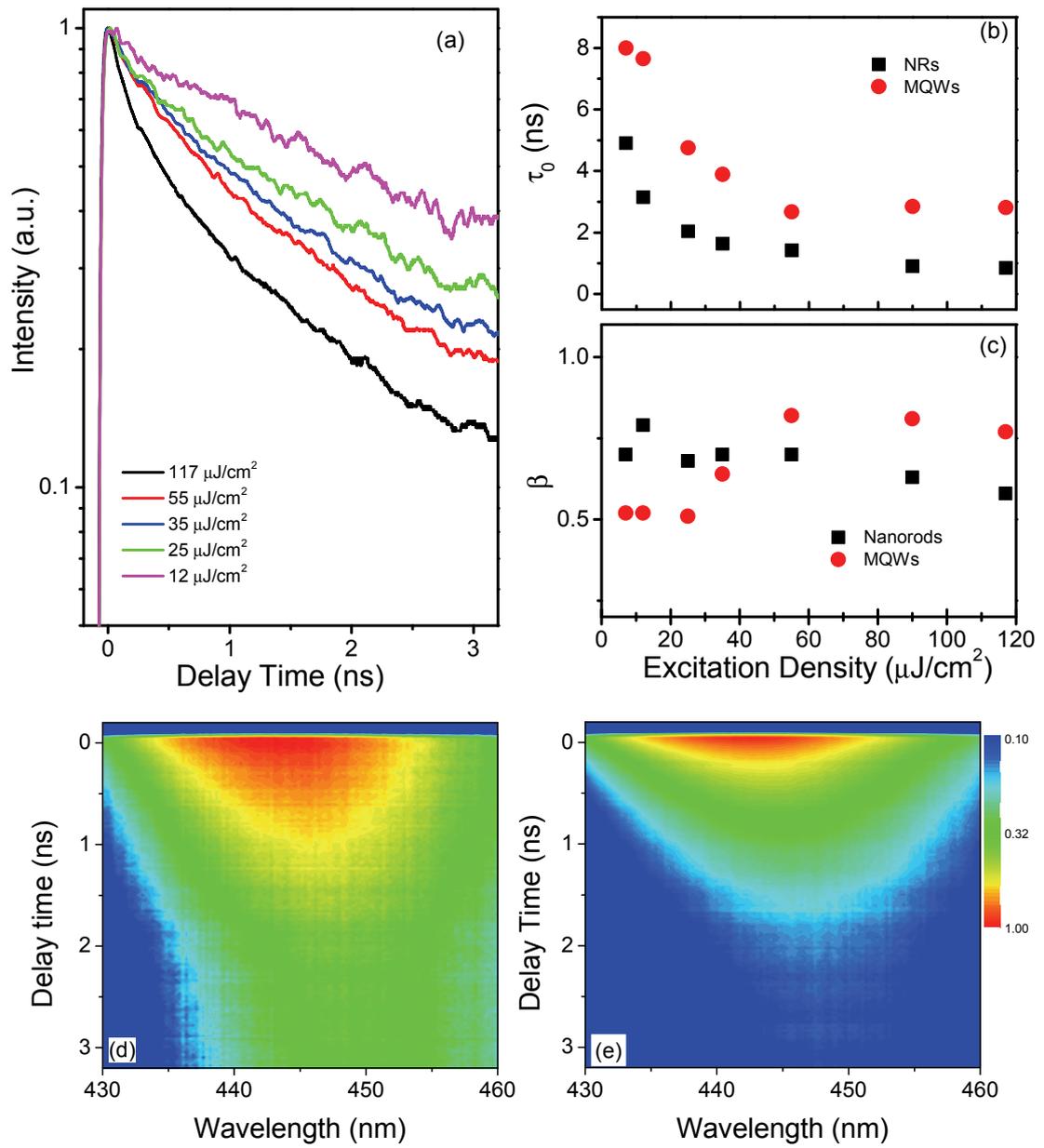

Jiang et al., Figure 3

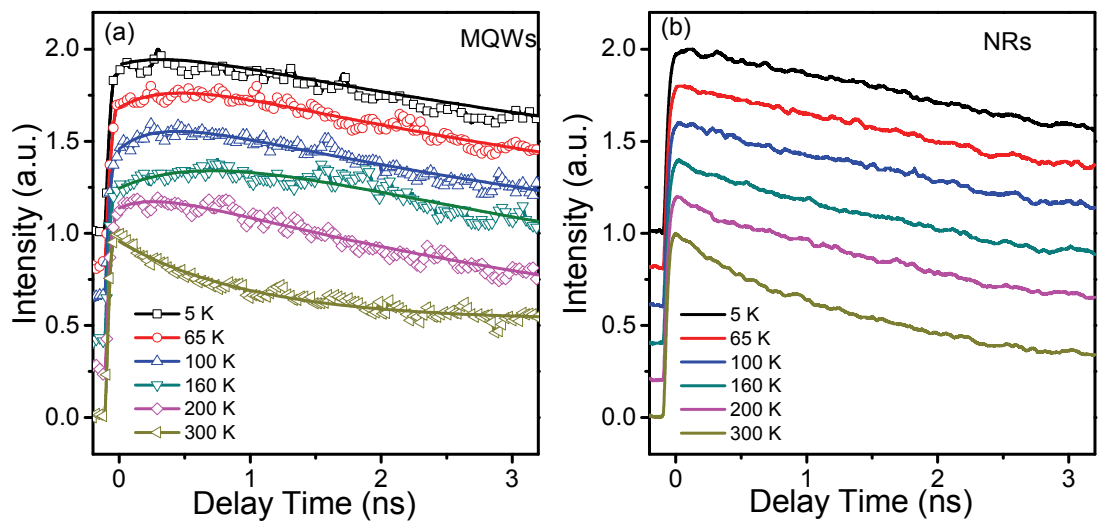

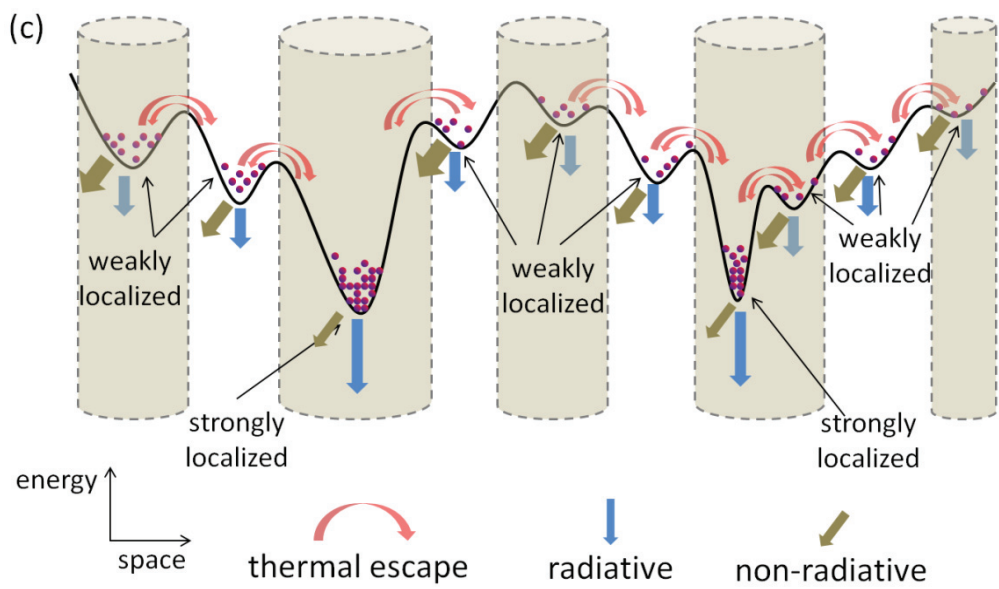

**Jiang et al., Figure 4**